\documentclass[conference]{IEEEtran}
\IEEEoverridecommandlockouts

\usepackage{graphicx}
\graphicspath{{Figures/}}
\DeclareGraphicsExtensions{.png,.jpg,.pdf}
\usepackage[normalem]{ulem} 
\newcommand{\li}{\uline{\hspace{0.5em}}}
\usepackage{enumitem}
\usepackage{url}

\usepackage{amsmath}
\usepackage{amsfonts}
\usepackage{amsthm}
\usepackage{float}
\usepackage{enumerate}
\usepackage{enumitem}
\usepackage{setspace}
\usepackage{balance}
\usepackage{xcolor}
\usepackage{color}
\usepackage{multicol}
\usepackage{multirow}
\usepackage{algpseudocode}
\usepackage[linesnumbered,ruled]{algorithm2e}
\SetKwInput{KwInput}{Input}
\SetKwInput{KwOutput}{Output}
\DeclareUnicodeCharacter{00A0}{ }
\usepackage[flushleft]{threeparttable}
\usepackage{listings}
\usepackage{textcomp}
\usepackage{tabulary}
\usepackage{booktabs}
\usepackage{upgreek}
\usepackage{subfigure}
\usepackage{overpic}
\usepackage{romannum}
\usepackage{tikz}

\newif\ifshort
\shortfalse

\title{Software-defined Dynamic 5G Network Slice Management for Industrial Internet of Things}

\author{
    \IEEEauthorblockN{Ziran Min\IEEEauthorrefmark{1}, Shashank Shekhar\IEEEauthorrefmark{2}, Charif Mahmoudi\IEEEauthorrefmark{2}, 
    Valerio Formicola\IEEEauthorrefmark{3}\footnote{work performed while at Siemens Technology.}, Swapna Gokhale\IEEEauthorrefmark{4}, Aniruddha Gokhale\IEEEauthorrefmark{1}}
    \IEEEauthorblockA{\IEEEauthorrefmark{1}Dept. of CS, 
Vanderbilt University,
Nashville, TN, USA.  Email:
    \{ziran.min,  a.gokhale\}@vanderbilt.edu}
    \IEEEauthorblockA{\IEEEauthorrefmark{2}Siemens Technology, Princeton, NJ 08540, USA.
  Email: \{shashankshekhar, charif.mahmoudi\}@siemens.com}
 \IEEEauthorblockA{\IEEEauthorrefmark{3}
  Dept. of ECE,
  California Polytechnic in Pomona,
  Pomona, CA, USA.
Email: vformicola@cpp.edu}
    \IEEEauthorblockA{\IEEEauthorrefmark{4}Dept. of CSE,
University of Connecticut,
Storrs, CT, USA.
  Email: swapna.gokhale@uconn.edu}}
\begin{document}
\maketitle
 \textcolor{red}{A version of this paper is under review by NCA 2022.}

\begin{abstract}
This paper addresses the challenges of delivering fine-grained Quality of
Service (QoS) and communication determinism over 5G wireless networks for
real-time and autonomous needs of Industrial Internet of Things (IIoT)
applications while effectively sharing network resources. Specifically, this
work presents DANSM, a software-defined, dynamic and autonomous network slice 
management middleware for 5G-based IIoT use cases, such as adaptive robotic repair.
\ifshort  
The novelty of our approach lies in (1) its heuristics algorithm that
solves an optimization problem at runtime to manage the network slices among
contending sub-tasks of the IIoT application, and (2) its realization as a
microservice that can be deployed in the 5G core control plane. 
\else     
The novelty of our approach lies in (1) the use of multiple M/M/1 queues to
formulate a 5G network resource scheduling optimization problem comprising 
service-level and system-level objectives; (2) the design of a heuristics-based
solution to overcome the NP-hard properties of this optimization problem, and
(3) the implementation of a software-defined solution that incorporates the 
heuristics to dynamically and autonomously provision and manage 5G network
slices that deliver predictable communications to IIoT use cases.
\fi       
Empirical studies evaluating DANSM on our testbed comprising a
Free5GC-based core and UERANSIM-based simulations reveal that the software-defined
DANSM solution can efficiently balance the traffic load in the data plane thereby 
reducing the end-to-end response time and improve the service performance by completing 
$34\%$ more subtasks than a Modified Greedy Algorithm (MGA), $64\%$ more subtasks than 
First Fit Descending (FFD) and $22\%$ more subtasks than Best Fit 
Descending (BFD) approaches all while minimizing operational costs.   
\end{abstract}


\ifshort
\vspace{-1mm}
\else
\fi
\section{Introduction}
\label{sec:Introduction}

The Fourth Industrial Revolution (Industry 4.0) is transforming a range of
today's industry verticals by bringing significant automation using modern
technologies, such as machine learning, real-time data processing and a gamut
of novel sensors, instruments, devices, hardware, software and networking.
This ecosystem that delivers the goals of Industry 4.0 like
automation, safety, timeliness, reliability and resilience is termed as
the Industrial Internet of Things
(IIoT)~\cite{corcoran2016mobile,narayanan2020key}. 

As an example, consider an IIoT use case of Adaptive Robotic Repair.  With the
recent disruptions in the industrial supply chain, it is increasingly becoming
important that the factories of today operate with zero human intervention
onsite and move towards the Lights-out Factory vision~\cite{lee2018total}.
Robotic arms are widely used to repair high-value components, such as shafts,
pistons, blades and molds. However, due to limited  compute resources available
on the robotic equipment and the need to work collaboratively, reliable and
real-time networking is of paramount importance.  

Wireless networks, such as Fifth-generation (5G) wireless, are attractive in
industrial environments as they enable mobility, eliminate the need for
expensive wiring needed by wired networks and overcome the hazards posed by
wired networks.  5G in particular supports~\cite{hassan2019edge} (a)
multiple base stations (gNB) that improve the signal strength and offer a
stable network connection, (b) mmWave and Multiple-Input Multiple-Output (MIMO)
technologies that enable electromagnetic waves to carry more raw data thereby
increasing network bandwidth and improving network latency, and (c) Network
Slicing (NS) that allows network providers to dynamically and
efficiently allocate network resources and offer differentiated services.   

The above-mentioned remote robotic repair IIoT use case consists of different
sub-tasks, such as workpiece scanning, defect
detection, tool path generation, workpiece milling, and milling
monitoring each with different priorities~\cite{alattas2019evolutionary}.  
The network requirements of the different involved sub-tasks are varied, which creates  
challenges to providing real-time packet inspection, delivering a high-level of 
Quality of Service (QoS) and generating an accurate usage report for every sub-task.  
This IIoT use case illustrates a  multi device and multi sub-task 
architecture. A 5G-enabled adaptive robotic repair IIoT system will 
require multiple co-existing network slices to deliver a real-time network solution. 
Moreover, considering the topology relationship among all sub-tasks and the different 
network resource consumption of each sub-tasks, the network resources that are assigned to
multiple network slices must be based on sub-task priority. Further, the heavy
network traffic generated by the sub-tasks will inevitably lead to queue buildup 
on the network data plane.

Previous studies have shown that high queuing delay within a 5G network leads
to rate variability~\cite{wang2019integrating} and adversely affects
QoS~\cite{irazabal2019active}. To address the aforementioned needs and
challenges in applying multiple network slices to the IIoT scenarios, we
present a software-defined approach called Dynamic and Autonomous Network Slice 
Management (DANSM). Since the 5G core by design separates the control plane from 
the data plane offering different functionalities as cloud-native microservices,
DANSM can easily be offered as an additional control plane microservice.  To that end
this paper makes the following contributions:

\ifshort  
\begin{itemize}
\item We describe DANSM's runtime heuristics algorithm that solves an optimization 
problem to autonomously assign the robotic repair sub-tasks to network slices;
\item We show how DANSM can be realized as a microservices middleware component
running in the 5G control plane; and
\item We show results of empirically validating our claims.
\end{itemize}

\else     

\begin{itemize}
  \item To efficiently utilize the network resources and improve network
  scalability, we present a topology sorting algorithm to decide the sub-task
  priority that determines the dynamic and autonomous assignment/recycling of
  network resources within each network slice. 
  \item To balance the load and minimize the queuing latency on the data plane,
  we present a multiple M/M/1 queue model of the data plane traffic and propose
  a heuristic algorithm to schedule the sub-tasks and dynamically manage the network
  resources based on the sub-task priority. 
 \item We show how DANSM provides sub-tasks with a network slice that helps to
  maintain and improve network services and requirements for a specific type of 
  sub-task thereby improving productivity in the industrial use case.  
  \item We show extensive empirical results evaluating our ideas.
\end{itemize}
\fi       

The rest of this paper is organized as follows:
Related work is briefly summarized in Section~\ref{sec:RelatedWork};
Section~\ref{sec:methodology} presents details of our approach;
Performance evaluations are presented and analyzed in Section~\ref{sec:Evaluation};
and finally, Section~\ref{sec:ConclusionAndFutureWork}  offers concluding
remarks alluding to future work.

\ifshort
\vspace{-1mm}
\else
\fi
\section{Background and Related Work}
\label{sec:RelatedWork}
This section provides background on 5G and then describes related research 
on dynamic management of network slicing, which is relevant to this research. 

\vspace{-1mm}
\subsection{Overview of 5G Wireless Networking}
\label{sec:5g}
The $5^{th}$ generation wireless networking is the latest cellular technology
that is being deployed around the globe. The 5G technology is designed to be
inherently cloud-native so that its functionality can be deployed in the form
of containerized microservices that can be managed and autoscaled by
frameworks, such as Kubernetes.  In its basic form, 5G comprises edge devices,
such as smart phones, called the \textit{User Equipment (UE).}  UEs communicate
with a base station called \textit{gNodeB} via a \textit{radio access network
  (RAN).} The core functionality of 5G that manages the user sessions,
authentication, network slicing, user packet forwarding and several other
important functions are realized as microservices and are part of what is
called the \textit{5G Core.}  While most of the capabilities, such as session
management, resource management and user authentication are control plane
responsibilities, the primary data plane function of routing and forwarding
user packets is carried out by the \textit{User Plane Function (UPF).}  5G's
\textit{Multi-access Edge Computing (MEC)} provides edge computing capabilities
to applications.

Further, the data plane within the 5G Core Network comprises multiple network
slices.  5G network slicing enables multiplexing of virtualized and independent
logical networks on top of common physical infrastructure. Presently, 5G
network slicing is categorized into 3 types: Enhanced Mobile Broadband (eMBB)
used by applications requiring ultra high bandwidth, Massive Machine-Type
Communications (mMTC) used in fast and energy-efficient communications, and
Ultra-reliable Low-Latency Communications (URLLC) used by applications needing
ultra-low latencies and reliable communications. By using different network
slices, we can satisfy the differentiated network requirements of IIoT
scenarios.  Note that 5G provides only the mechanisms but algorithms are needed
to effectively manage these slices. 

\vspace{-1mm}
\subsection{Comparison with Prior Work}
\label{sec:prior}
Dynamic network slicing technology, which virtualizes shared physical networks
by providing multiple network services, is widely applied in both academic and
industrial areas.  For example, Xiao \textit{et al.}~\cite{xiao2018dynamic}
proposed the concept of dynamic network slicing.  They developed an overlapping
coalition-formation game to investigate the distributed cooperation and joint
network slicing between fog nodes while considering traffic variation. Their
results show that their architecture can significantly maximize
utilization while balancing the workloads on fog
nodes. In~\cite{feng2020dynamic}, the authors proposed a dynamic network
slicing and resource allocation approach to investigate the operator's revenue
escalation problem under dynamic traffic in a mobile edge computing
system. This approach optimizes the network slice admission in the long term
and resource allocation in the short term. However, their approach considers
only the transmission delay while ignoring the queuing latency, which will
increase the end-to-end latency, thereby affecting system performance.

To intelligently assign and redistribute resources among multiple tenants, Raza
\textit{et al.}~\cite{raza2018dynamic} leverage 5G orchestration
functionalities to support dynamic network slicing, which jointly provisions 
the network resources from different domains, such as
radio, transport, and cloud.  They formulated a mixed-integer linear
programming problem and designed a heuristic to solve it. Their evaluations
show that dynamic slicing can improve the virtual network rejection probability
by more than one order of magnitude. Like prior works, our approach can achieve
the system-level objective of balancing the data plane traffic load among
different network slices.  Additionally, we also achieve the service-level
objective of minimizing both the queuing and transmission time.  

The work in~\cite{raza2018dynamic} proposes a  dynamic slicing approach to
assign and redistribute resources among multiple tenants over 5G networks, where
the 5G core network is treated as a black box and its details ignored. 
In their simulation, the authors focused on the control
plane complexity and considered only the tenant's requests.  
In contrast, our approach focuses on both the control plane and data plane
requests.  Moreover, we also consider the role of 5G core functions and
utilize  multiple M/M/1 queues to schedule the packets from the User Equipment (UE) to User Plane Function (UPF), 
which is the primary data plane function in 5G. 

In summary, our work utilizes a software-defined, 5G-based dynamic slicing approach 
to allocate network resources for the IIoT usecases with different sub-tasks 
having different priorities, while satisfying the real-time and high throughput
requirements. Compared to previous dynamic slicing-based approaches, our work
applies the M/M/1 queuing theory to model the network traffic and formulates an
optimization problem. Moreover, we are able to improve the
utilization of network resources, significantly reduce both the queuing latency
and the transmission time, and effectively balance the load among different
network slices all at once.

\ifshort
\vspace{-1mm}
\else
\fi
\section{Methodology}
\label{sec:methodology}

\begin{figure}[htb]
\vspace{-2mm}
\centering
\includegraphics[width=0.5\textwidth,keepaspectratio]{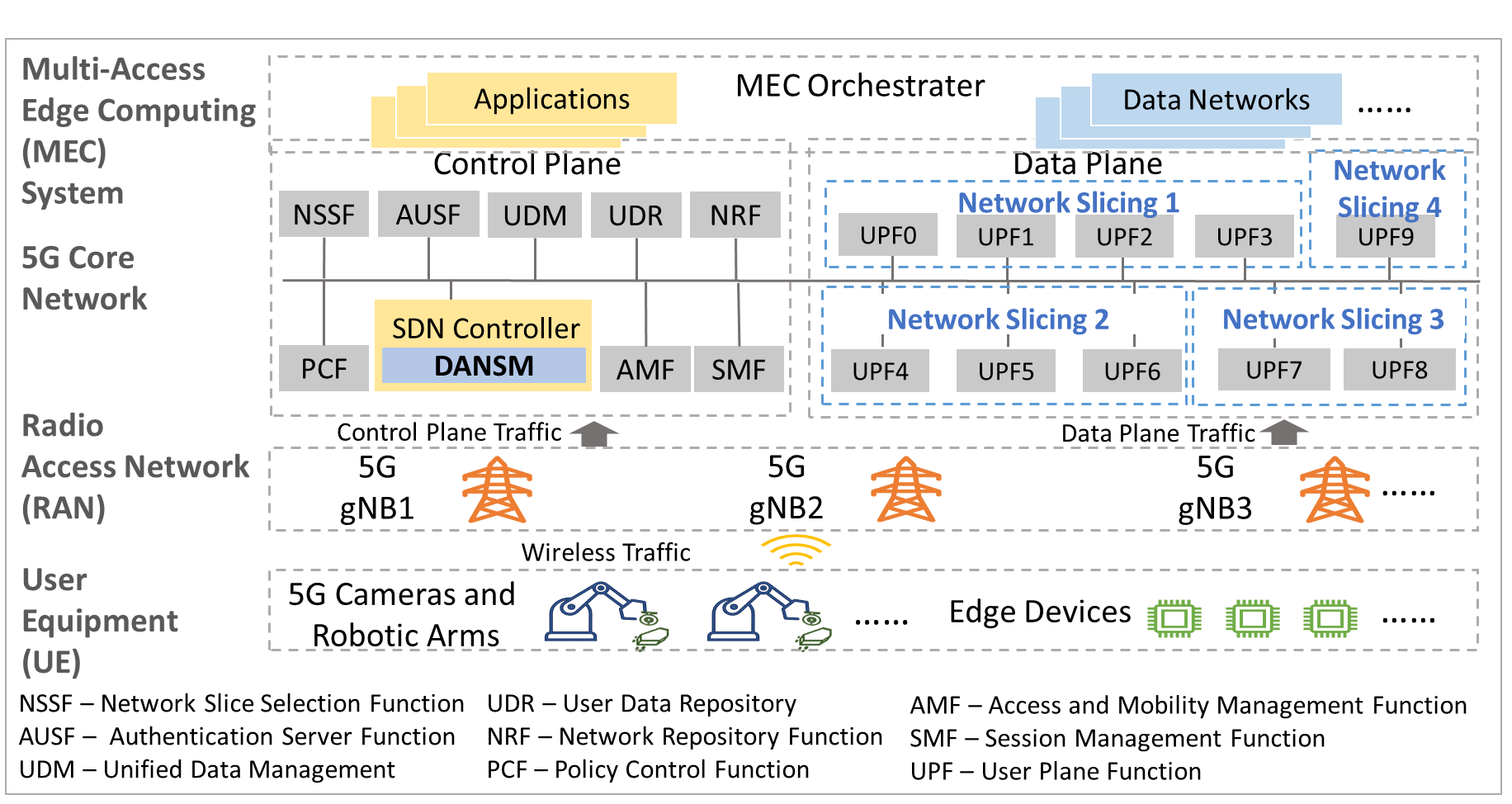}
\vspace{-2mm}
\caption{System Architecture}
\label{fig:SystemArchitecture}
\vspace{-2mm}
\end{figure}

This section describes DANSM, which is designed to operate in a 5G
ecosystem shown in Figure~\ref{fig:SystemArchitecture}.  We first
provide details on a concrete industrial use case to describe the research.
\ifshort  
Finally, we describe our approach focusing on the runtime heuristics;~\cite{Min2021DANSM}
provides details on the optimization problem formulation.\footnote{Due to space
  constraints, this paper focuses only on the runtime aspects leaving the design-time
  details to a technical report.} 
\else     
Finally, we describe the DANSM approach in detail.
\fi       

\vspace{-1mm}
\subsection{IIoT Case Study and Key Issues}

Our industrial IIoT use case is a 5G-based adaptive robotic repair system that
we use to motivate and describe our research. The UEs\footnote{See Section~\ref{sec:5g} for
background on 5G and terminology.} in our use case comprise
cameras, robot arms and edge devices, which are assigned to different repair
tasks each of which generates various types and volumes of data.  An adaptive robotic
repair task consists of 5 sub-tasks: workpiece scanning, defect detection,
tool-path generation, robot milling and milling monitoring.
Note that our industrial IIoT use case serves only as a driving example to describe and evaluate 
our research; the DANSM research described here is broadly applicable beyond this case study.
\ifshort  
\else     

\fi       

As part of the adaptive robotic repair workflow, a 5G camera first scans the
target workpiece and records the key features. Then, the recorded information
(image or video) is sent from the 5G camera to the 5G MEC devices.  The MEC
devices, which include sensors, actuators, and other endpoints, will also
collaborate to help detect any defects. Then the MEC will compare the scanned
and standard workpiece features and generate a comprehensive \textit{repairing
  tool path} that is sent to the robotic arm.  On receiving the 
repairing tool path, the robotic arm will start milling the target
workpiece. The 5G camera will also monitor the milling process to prevent
accidents. 

Every sub-task within an adaptive robotic repair is associated with a network
slice, which consists of a number of UPFs. The number of UPFs within a network
slice is dependent on the sub-task's priority. The UPFs of the same
network slice share the same network configuration, such as bandwidth, which
guarantees that the packets from UEs routed in the same network slice but
different UPFs will be processed by the same network service. The UPFs of
different network slices have different network configurations, which provides
differentiated network services for different sub-tasks.  Moreover, a UPF routes
packets not only to the UE but also to the data networks within the MEC system,
which is responsible for providing time-sensitive and compute edge
services, such as the tool path generation sub-task within an adaptive robotic
repair.   

The varied data volumes from the adaptive repair devices,
however, can lead to network congestion, causing a number of queues to build
up, particularly in the data plane, thereby increasing the queuing and service
times of the adaptive repair system, and hence adversely impacting the performance of
the adaptive repair. Further, if an \emph{ad hoc} task-to-robot assignment policy
is used, then the adaptive  repair device that completes its assigned sub-task
earlier will simply wait until being assigned a new sub-task thereby wasting
precious resources, which can prolong the overall completion time of the entire
repair thereby hurting manufacturing productivity.

\ifshort  
\vspace{-1mm}
\subsection{DANSM Approach}
\vspace{-1mm}
DANSM is responsible for defining the priorities of the different sub-tasks of
an adaptive repair based on which it can determine the number of UPFs and
network slices needed to satisfy the QoS needs.  Moreover, it must minimize any
congestion and hence queuing, and deliver the latency requirements of the
different sub-tasks of the repair process.  To that end, it must dynamically
and autonomously manage the network slices that are allocated to the different
sub-tasks.
Figure~\ref{fig:AlgorithmFlowchart} depicts DANSM's heuristics algorithm. 

\begin{figure}[htb]
\vspace{-2mm}
\centerline{\includegraphics[width=0.85\columnwidth,keepaspectratio]{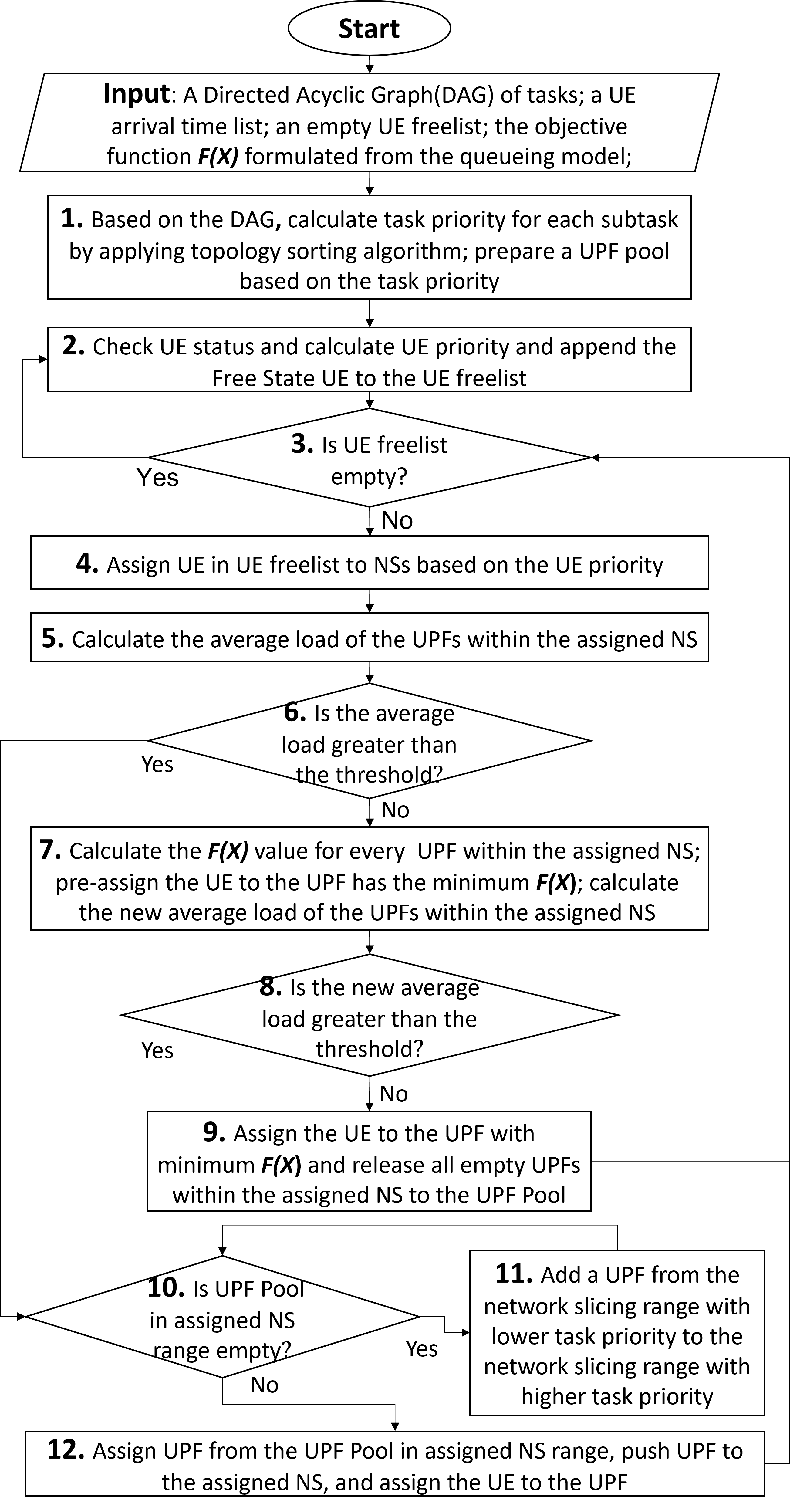}}
\vspace{-2mm}
\caption{Heuristic Network Slice Management Algorithm}
\label{fig:AlgorithmFlowchart}
 \end{figure}

\begin{figure*}[htb]
\vspace{-2mm}
\centerline{\includegraphics[width=0.85\textwidth,keepaspectratio]{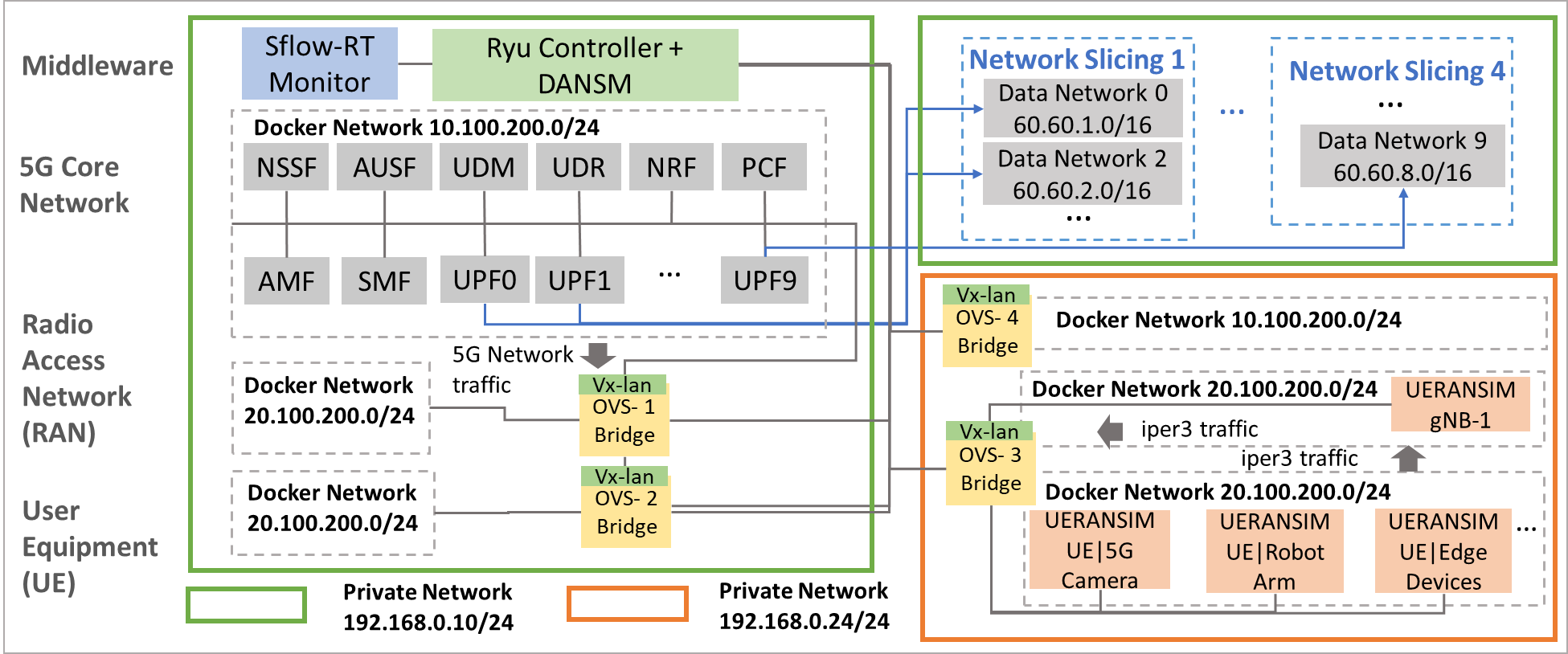}}
\vspace{-2mm}
\caption{Experiment Testbed (Two physical machines were used and indicated in green and orange boxes.)}
\label{fig:Testbed}
\end{figure*}

In~\cite{Min2021DANSM}, we describe a topological sorting approach that determines sub-task
priorities, and a multi-M/M/1 queuing model that is used to formulate a
mathemical optimization problem that meets the objectives of minimizing queuing
and latency.  Due to the NP-hard nature of the optimization problem, DANSM
defines a heuristics-based algorithm to make rapid runtime decisions on
dynamically managing the network slices, which is explained below.

\ifshort  

As shown in Figure~\ref{fig:AlgorithmFlowchart}, DANSM involves some feedback loops based on
which it makes adaptation decisions.  The algorithm accepts a directed acyclic
graph (DAG) of sub-tasks of the robotic repair task and their dependencies.  In
Step 1, sub-task priorities are computed by a topological sort as discussed in
our detailed report~\cite{Min2021DANSM}.  A pool of UPF components for the different
priorities is then created for routing UE traffic.  In Step 2, the algorithm
checks if a UE (i.e., a robotic arm) is free, in which case it is added to the
freelist, or still working.  Step 3 checks if there is any available UE that
will need a slice and is assigned a task.  If no UE is available, the algorithm
waits until a UE is available.  Note that initially all UEs are free.

Step 4 picks a UE from the freelist and assigns it a network slice with the
appropriate priority. Step 5 then determines the load on the UPFs belonging to
that slice. If this computed load is greater than some threshold (condition
check in Step 6), then it may indicate an overload situation. In an overload
situation, Step 10 checks if there are any available UPFs to handle traffic for
this overloaded slice. If none is available, then an available UPF from a lower
priority pool is temporarily elevated to higher priority and utilized for this
slice (Step 11). Otherwise, a UPF is chosen from the current UPF pool.

Step 7, which is executed in the non-overload case, computes the expected
latency for every UPF within the slice under consideration and assigns the
chosen UE from the freelist to the UPF that provides the best latency.  Step 8
once again checks for the overload condition.  If no overload, then the UE is
actually assigned the minimum latency UPF, and any unused UPFs are returned to
the UPF pool.

\else     
\vspace{-1mm}
\subsection{Heuristic Approach}
\label{sec:HeuristicApproach}

Our optimization problem formulation shown in Eq.(9) belongs to the class of
dynamic scheduling problems for multiple parallel servers/queues, which has
been shown to be NP-hard~\cite{kravchenko1998scheduling}.  Hence, to rapidly
solve our optimization problem at run-time as part of the dynamic and
autonomous approach, we propose a heuristic algorithm.  Our approach balances
the load on the UPF side, and minimizes queuing and propagation latency under
dynamic traffic conditions for real-time communication use cases in IIoT. 

\begin{figure}[htb]
\vspace{-2mm}
\centerline{\includegraphics[width=0.87\columnwidth,keepaspectratio]{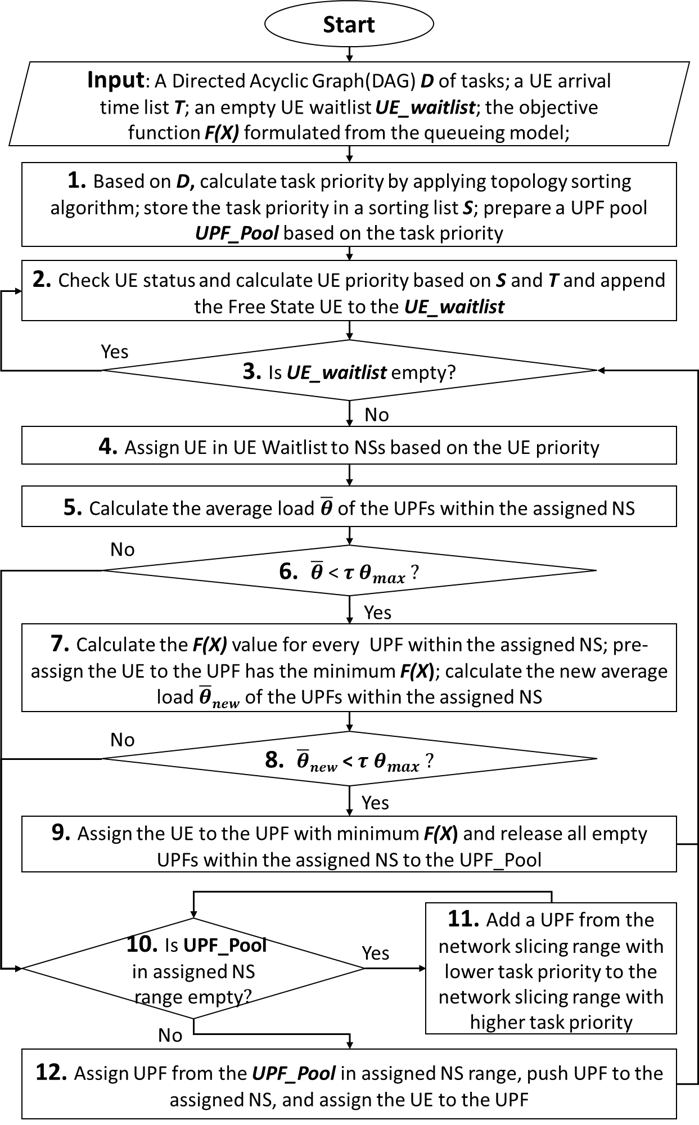}}
\caption{Heuristic Scheduling Algorithm Involving an Autonomous Feedback Loop}
\vspace{-2mm}
\label{fig:AlgorithmFlowchart}
\vspace{-2mm}
 \end{figure}
 
We use our adaptive robotic repair case study to present our heuristic
algorithm shown as a flowchart in Figure~\ref{fig:AlgorithmFlowchart}. The
input to our algorithm includes the following: (a) an adaptive robotic repair
Directed Acyclic Graph (DAG) $D$, which is  provided by the DANSM user and
which includes the topology relationship among all the repairing sub-tasks; (b)
a UE arrival time list $T$; (c) a UE waitlist $UE\_{waitlist}$, which is a
matrix and used for storing all the UE statuses and UE priorities; and (d) an
objective function $F(x)$, as shown in Eq.(9) formulated from the multiple
M/M/1 queuing model that utilizes the UE to UPF assignment information.  

The algorithm works as follows:
    In step 1, our algorithm will calculate the task priorities by
    utilizing $D$ and apply the Topology Sorting
    Algorithm~\cite{kahn1962topological} on $D$. Figure~\ref{fig:DAG} shows how
    to generate a DAG and calculate the task priorities for a multi sub-tasks
    IIoT usecase. Then, the algorithm will store the task priority in a list
    $S$ and create a UPF pool, which stores a number of available UPFs for all
    the network slices.  
    In step 2, the algorithm will check the UE status (either at
    initialization or as the system evolves over time) and calculate the UE
    priority based on the UE buffer status, task priority, and UE arrival
    time. The UE buffer status and the UE arrival time is obtained from the SDN
    controller. If the SDN controller detects 0 bytes in a UE buffer, we set
    the UE to the Free state, which indicates that the UE has finished its
    previous task and is waiting for a new task. Otherwise, the UE will be set
    to the Service state, which indicates that the UE is still working on the
    current task. 
    
    The algorithm will then update the UE status based on their
    buffer status. This step will run periodically. 
    All the free state UEs will be added to the $UE\_{waitlist}$. Each UE in
    the $UE\_{waitlist}$, at the specific time, is responsible for one sub-task
    within the adaptive robotic repair and will be assigned to the network
    slice (NS), which is matched with the sub-task, based on their UE priority.  
    In step 3, our algorithm will check if the $UE\_{waitlist}$ is empty
    or not. If it is empty, we go back to step 2 and check all the UE status
    again till the $UE\_{waitlist}$ is not empty, which means there is at least
    one UE that can be assigned the next sub-task.  
    Otherwise, we go to step 4, where for every available UE we assign it
    to the matched network slice based on their task priority and where network
    slices may also have different priority.  
    
    In step 5, the algorithm will calculate the average load of the
    UPFs within their assigned network slice. This computation is needed to
    assist in load balancing. 
    In step 6, we check if the average load is greater than the maximum
    load times the threshold coefficient $\tau$. If this is the case, we assume
    all the UPFs within the matched NS are at a risk of overload. If the
    current assigned NS is overloaded, then the algorithm will check if the
    $UPF\_{Pool}$ in the assigned NS range has any available UPFs in step
    10. If the assigned NS range of $UPF\_{Pool}$ runs out of UPFs, then the
    algorithm goes to step 11 and pulls a UPF from the NS range with lower
    priority and elevates it to the NS range with higher priority. Otherwise,
    the algorithm will go to step 12 and directly pull a UPF from the UPF pool
    in the assigned NS range (which could include the just elevated UPF), push
    it to the matched NS, and assign the UE to the UPF.  Thereafter, the flow
    goes back to Step 3 as described above.  
    
    On the other hand, if in step 6 the average load is less than the
    maximum load times $\tau$, the algorithm will go to step 7 and calculate
    the $F(x)$ value according to Eq.(9) for every UPF within the assigned NS
    and pre-assign the UE to the UPF, which has a minimum $F(x)$. "Pre-assign"
    means that the UE is temporally and logically assigned to the UPF for
    calculating the new average load $\overline\theta_{new}$.  
    Then, in step 8, the algorithm will check if $\overline\theta_{new}$
    is less than the maximum load times $\tau$. If yes, the algorithm will go
    to step 9 and the UE will be physically assigned the UPF, which has the
    minimum $F(x)$, and then go back to Step 3. Otherwise, we will go to step
    10 because this is a case where potentially all the UPFs within the
    assigned NS have the risk of overload, and hence will perform the same
    steps as described before for the overload case. 
In our algorithm, every UE in the $UE_waitlist$ is supposed to scan all the
UPFs within the matched network slice. Therefore, the runtime complexity of our
algorithm is non-linear and denoted by $O(nr)$, where $n$ indicates the number
of UE and $r$ indicates the number of UPFs. 
\fi       


\else     

\vspace{-1mm}
\subsection{Priority-based and Queuing-theoretic Modeling}
To address the unpredictable waiting time issue, we introduce the notion of
\emph{Task Priority} and \emph{UE Priority.} Task priority helps to dynamically
assign/recycle resources allocated to the network slices, which are associated
with the repair sub-tasks, while UE priority ensures that the UE that has
completed its task early, can be assigned a new task at the earliest. The Task
Priority can be calculated by applying the Topology Sorting
Algorithm~\cite{kahn1962topological} on the task flowchart, which represents
the topology relationship among repair sub-tasks as shown in
Figure~\ref{fig:DAG}. The task that executes early in the task flowchart will
get higher priority. The network slice with higher task priority will be
assigned more network resources initially. The UE Priority is formulated based
on the next task priority and the current task start time, which is also
referred to as the UE arrival time, and is represented as $UE Priority = Task
Priority + \frac{1}{Arrival Time}$. The UE, whose new task has higher priority,
will be assigned to the matched network slice early and has more choices when
choosing the UPFs within the matched network slice. Therefore, those UEs will
have a higher chance of avoiding the overloaded UPFs. 

\begin{figure}[htb]
\vspace{-2mm}
\centerline{
\includegraphics[width=8
cm, height=4cm]{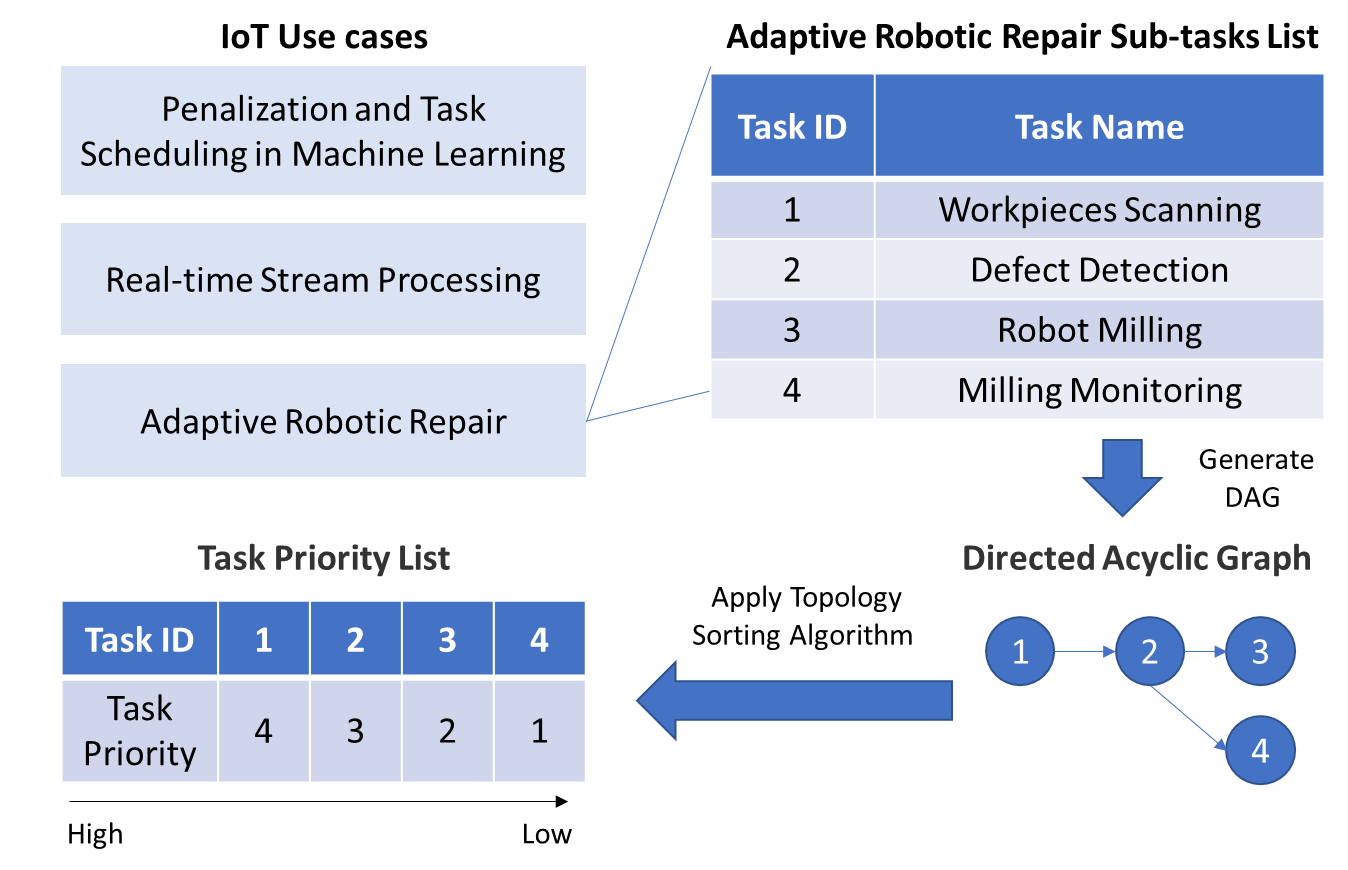}}
\vspace{-2mm}
\caption{DAG Flowchart}
\label{fig:DAG}
\vspace{-2mm}
 \end{figure}

Prior studies have adopted the M/M/1 and M/M/N queuing models to schedule IoT
network packets in both wired~\cite{9659533,deva2020efficient} and wireless
networks~\cite{roy2021overview}.  Additionally, studies have shown that both
wired~\cite{metzger2019modeling} and wireless~\cite{keeler2018wireless} traffic
follow Poisson distribution in IoT scenarios.  Since the robotic repair task is an
IIoT usecase and backed by success of prior efforts, we use the M/M/1 model to schedule
the traffic in the data plane. Specifically, our work uses the M/M/1 queuing
model to model the latency, service time and transmission delay of requests in
the 5G UPF. In our system, the flows representing arrival
requests are assumed to be from independent UEs with the interval of arrival
time and the server processing time constituting a negative exponential
distribution. Therefore, requests from UEs to UPFs follow a Poisson
distribution thereby justifying the use of M/M/1 modeling.  Moreover,
considering the heterogeneous traffic from UEs and the topology
relationship among robotic adaptive repair sub-tasks, we provision a pool of
UPFs and configure multiple network slices allocating/deallocating resources
dynamically to handle and adapt to the traffic from the UE side.  Every sub-task
is assigned to a specified network slice, and the resource\footnote{A resource 
corresponds to the number of UPFs.} within the network slice is dynamically changed 
based on sub-task priority and real-time UE request rates. These decisions are
made by DANSM software-defined design operating from the 5G control plane. 


\vspace{-1mm}
\subsection{Optimization Problem Formulation}
\label{sec:ProblemFormulation}

Our system queuing model is shown in Figure~\ref{fig:Queuing Model} while Table~\ref{tab:notations} shows the notations used in the problem formulation described in this section. 

\vspace{-2mm}
\begin{figure}[htb]
\centerline{
\includegraphics[width=8.5cm, height=2.2cm]{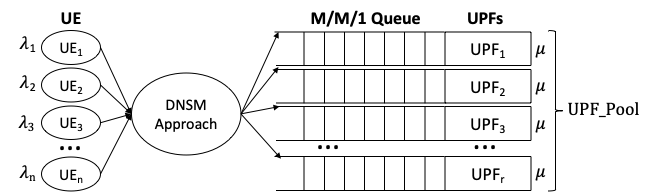}}
\vspace{-2mm}
\caption{System Queuing Model}
\label{fig:Queuing Model}
\vspace{-2mm}
\end{figure}

We assume there are $n$ UEs and the request rate of the $i^{th}$ UE is denoted by $\lambda_i$. The total UE request rate is $\sum_{i=1}^{n}\lambda_{i}$. We assume there are $s$ network slices and $r$ UPFs in the $UPF\li Pool$. The initial number of UPFs within every network slice is decided by the coefficient $\alpha$ and the $Task\  Priority$. The $\alpha$ is provided by the users and can decide the initial size of each network slice.

\newcommand{\tabincell}[2]{\begin{tabular}{@{}#1@{}}#2\end{tabular}}
\begin{table}[]
  \centering
  \small
  \vspace{-2mm}
  \caption{LIST OF VARIABLES}
  \label{tab:notations}
  \begin{tabular}{lp{6.3cm}}
    \\[-2mm]
    \hline
    \hline\\[-2mm]
    {\bf Symbol}&\qquad {\bf Meaning}\\
    \toprule
    \vspace{1mm}\\[-3mm]

    \vspace{0.1mm}
    $n$      &   \tabincell{l}{ the number of User Equipments (UEs).} \\
      \toprule
      
    \vspace{0.1mm}
    $S$      &   \tabincell{l}{ Task priority list $S$ = \{($task_{1}$, $tp_{1}$), ($task_{2}$, $tp_{2}$),\\ ($task_{3}$, $tp_{3}$), \dots, ($task_{s}$, $tp_{s}$) \}; total $s$ tasks,\\ which are mapped to $s$ network slices; $task_{s}$ is\\ the id of $s^{th}$ task and $s^{th}$ network slice; $tp_{s}$ is \\the task priority of $s^{th}$ task.} \\
      \toprule
      
    \vspace{0.1mm}
    $UPF\li Pool$      &   \tabincell{l}{ A provisioned UPF pool $UPF\li Pool$ = \{$upf_{1}$,\\ $upf_{2}$, $upf_{3}$, \dots, $upf_{r}$\}; total $r$ UPFs; $upf_{r}$ is \\the id of $r^{th}$ UPF. }\\
      \toprule

    \vspace{0.1mm}
    $\alpha$      &   \tabincell{l}{The coefficient of the number of UPFs in\\ $UPF\li Pool$; $\alpha\sum_{k=1}^{s}tp_{k} = r$.  In the \\$UPF\li Pool$, there are $\alpha tp_{1}$ UPFs 
    prepared for \\the $1^{st}$ network slice; the range of UPFs within \\the $1^{st}$ network slice in the $UPF\li Pool$ is [$1$,\\ $\alpha$$tp_{1}$]. There are $\alpha$$tp_{s}$ UPFs prepared for the $s^{th}$\\ network slice and the range of $s^{th}$  network slice\\ in the $UPF\li Pool$ is [$\alpha$$\sum_{k=1}^{s-1}tp_{k}$, $\alpha$$\sum_{k=1}^{s}tp_{k}$]. }\\
      \toprule
    \vspace{0.1mm}
    $\lambda_{i}$    &	\tabincell{l}{the request rate of $i^{th}$ UE. }\\
      \toprule
      
     \vspace{0.1mm}
    $\mu$          &  \tabincell{l}{the service rate of all UPF.}\\
      \toprule
      
     \vspace{0.1mm}
    $X_{k}$ &  \tabincell{l}{the UE requests to UPF  assignment matrix of \\$k_{th}$ network slice; $x_{ij}=1$ means that $i^{th}$ UE\\ send requests to $j^{th}$ UPF , $x_{ij}\in\{0,1\}$, $\forall{i,j}$. } \\
      \toprule
    \vspace{0.1mm}
    $\theta_j$ & the load of $j^{th}$ UPF. \\
      \toprule
    
    \vspace{0.1mm}
    $\overline{\theta_{k}}$ &\tabincell{l}{the average load of all the UPFs within $k^{th}$ \\network slice.} \\
      \toprule
      
     
    
    \vspace{0.1mm}
    $l_{i,j}$ &\tabincell{l}{ the length of packet from $i^{th}$ UE to $j^{th}$ UPF. }\\
      \toprule
      
    \vspace{0.1mm}
    $d_{ij}$ & the transmission rate between $i^{th}$ UE and $j^{th}$ UPF. \\
      \toprule
      
    \vspace{0.1mm}
    $w_{1},w_{2}$ &\tabincell{l}{ weight factors, which will be tuned accordingly.}\\
    \hline
    \hline
  \end{tabular}
\end{table}

In the UPF Pool, there are $\alpha tp_{1}$ UPFs prepared for the $1^{st}$ network slice and the range of UPFs within the $1^{st}$ network slice in the $UPF\li Pool$ is [$1$, $\alpha$$tp_{1}$]. We assume $\alpha = 1$ , $r=10$ and use the example in Figure~\ref{fig:DAG}. Then, the  $1^{st}$ network slice will have $\alpha tp_{1} = 1*4=4$ UPFs, and the range of $1^{st}$ network slice in the $UPF\li Pool$ is [$1$, $4$]. And the $2^{nd}$ network slice will have $\alpha tp_{2} = 1*3=3$, the range of $2^{nd}$ network slice is [$5$, $7$]. The total number of UPFs within the 4 network slices will be $4+3+2+1=10$. 
There are $\alpha$$tp_{s}$ UPFs prepared for the $s^{th}$ network slicing and the range of $s^{th}$  network slicing in the $UPF\li Pool$ is [$\alpha$$\sum_{k=1}^{s-1}tp_{k}$, $\alpha$$\sum_{k=1}^{s}tp_{k}$], thus $\alpha\sum_{k=1}^{s}tp_{k} = r$.  Every UPF has a queue, thus there are r M/M/1 queues in our adaptive repair system. Our DANSM algorithm will assign the incoming requests from $n$ UEs to the $r$ UPF queues. Then, we assume all the UPFs have the same service rate, which is denoted by $\mu$. The UE to UPF assignment is stored in the binary matrix $X$. The matrix element $x_{ij}$ is 1, if the $i^{th}$ UE is assigned to $j^{th}$ UPF, otherwise  $x_{ij}$ is 0.

Assuming that all the UE request arrivals are  mutually independent and follow a Poisson process. Therefore, the load on the system is  is represented by $\sum_{i=1}^{n}\lambda_{i}$. The total load should be less than $r\mu$ for a stable system. 
The load of the $j^{th}$ UPF is represented by:
\begin{equation}
\vspace{-1mm}
\small
\begin{aligned}
     \theta_j = \sum_{i=1}^{n}\lambda_{i}x_{ij}
\end{aligned}
\vspace{-1mm}
\end{equation}
and the average load among all UPF within the $k^{th}$ network slice is represented by:
\begin{equation}
\vspace{-1mm}
\small
\overline{\theta_{k}} = \frac{1}{\alpha tp_{k}} \sum_{j=\alpha tp_{k-1}}^{\alpha tp_{k}} \theta_j
\vspace{-1mm}
\end{equation}
Applying Little's law, the expected queuing time on the $j^{th}$ UPF before a request from an UE is served can be represented by:  
\begin{equation}
\vspace{-1mm}
\small
\begin{aligned}
     W_{q} =\frac{\theta_{j}}{\mu(\mu-\theta_{j})}
\end{aligned}
\vspace{-1mm}
\end{equation}
and the expected end-to-end response time between an UE and $j^{th}$ UPF, which is the sum of the request
queuing time and the UPF service time, is represented by:
\begin{equation}
\vspace{-1mm}
\small
\begin{aligned}
     W_{s} = W_{q} + \frac{1}{\mu} = \frac{1}{\mu-\theta_{j}} 
\end{aligned}
\vspace{-1mm}
\end{equation}
The transmission time is decided by the length of the packet and the transmission rate between $i^{th}$ UE and $j^{th}$ UPF. Considering the different types of UEs and the different sub-tasks,  the lengths of packets generated by the UE are different. Moreover, considering the different UPFs configured for different network slices, the network configuration of UPFs, such as bandwidth, are different. Therefore, the transmission time varies and can be represented by:
\begin{equation}
\vspace{-1mm}
\small
\begin{aligned}
     W_{t} =\sum_{i=1}^{n}\frac{l_{i,j}}{d_{ij}}
\end{aligned}
\vspace{-1mm}
\end{equation}
The overall latency between $i^{th}$ UE and $j^{th}$ UPF, which is the sum of queuing latency, the UPF service time and transmission time, is represented by:
\begin{equation}
\vspace{-1mm}
\small
\begin{aligned}
     W_{s} + W_{t} = \frac{1}{\mu-\theta_{j}} + \frac{l_{i,j}}{d_{ij}}
\end{aligned}
\vspace{-1mm}
\end{equation}
The average queuing latency and the transmission time of all the UPFs within the $k^{th}$ network slicing is represented by:
\begin{align}
\vspace{-1mm}
\small
\quad G(X_{k}) =\frac{1}{n}( \sum_{i=1}^{n} \sum_{j=\alpha tp_{k-1}}^{\alpha tp_{k}} \frac{x_{ij}}{\mu-\theta_{j}}+\frac{l_{i,j}}{d_{ij}})
\vspace{-1mm}
\end{align}
To balance the load of UPFs and minimize overall latency for every UPF, we also formulated the variance of latency to avoid the extreme case, where all the UE loads are assigned to a few UPFs. The variance of the queuing latency and the transmission time of all the UPFs within the $k^{th}$ network slicing is represented by:
\begin{align}
\vspace{-1mm}
\small
\quad V(X_{k}) =\frac{1}{n}( \sum_{i=1}^{n} \sum_{j=\alpha tp_{k-1}}^{\alpha tp_{k}} (\frac{x_{ij}}{\mu-\theta_{j}}+\frac{l_{i,j}}{d_{ij}}-G(X)))^{2}
\vspace{-1mm}
\end{align}
In this work, we aim to minimize both the mean and the variance of  queuing latency and the transmission time. Considering the difference of magnitude between the mean and the variance value, we tuned $w_{1}$ and $w_{2}$ accordingly as weight factors. The following illustrates the problem formulation:  
\begin{align}
\vspace{-1mm}
\small
\text{min}\quad F(X) = \sum_{k=1}^{s}(w_{1}G(X_{k})+w_{2}V(X_{k}))
\vspace{-1mm}
\end{align}
\begin{align}
\vspace{-1mm}
\small
\text{s.t.} \quad &\sum_{i=1}^{n}\lambda_{i}x_{ij} < \mu, \forall{j}\\
&\sum_{i=1}^{m}x_{ij} = 1, \forall{i}\\
& x_{ij} \in\{0,1\}, \forall{i,j}\\
&\alpha\sum_{k=1}^{s}tp_{k}= r 
\vspace{-1mm}
\end{align}
Eq.~(9) aims to minimize both the average UE queuing latency and the average transmission time in standalone 5G network.  Eqs.~(10)-(13) represent the constraints.  As mentioned earlier: a) for each UPF, the sum of the request rates from all connected UEs should be less than the UPF's processing rate $\mu$; b) each UE can only be connected to one UPF at a time; c) the UE connection decision is encoded in a matrix $x_{ij}$ with binary elements; d) the number of UPFs is limited and the task has higher priority will have more prepared UPFs in the UPF pool.


\fi       

\vspace{-1mm}
\subsection{DANSM Implementation}
DANSM provides software-defined resource management for 5G network slices.
It is realized as a microservices component that can be deployed in the control
plane of the 5G core along with other components.  The 5G architecture makes
this design choice easy to implement without any invasive changes to existing
components.  DANSM is implemented in Python.\footnote{DANSM is available in open-source
from \url{https://github.com/minziran/DANSM}.}


\section{Empirical Evaluation} 
\label{sec:Evaluation}
This section reports on the results of extensive evaluations we conducted validating our proposed DANSM approach.  


\ifshort 

\else  
\begin{figure}[htb]
\vspace{-2mm}
\centerline{\includegraphics[width=0.5\textwidth,keepaspectratio]{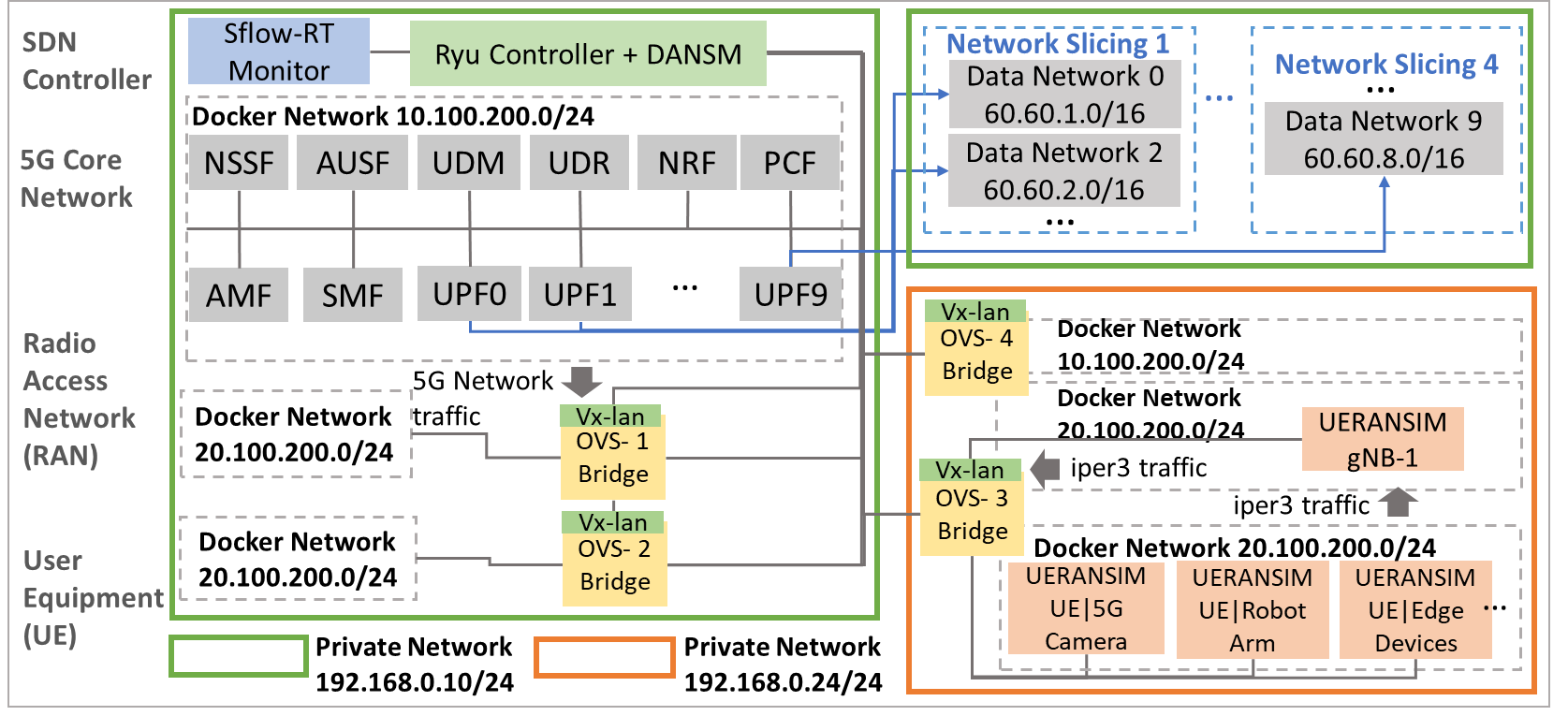}}
\vspace{-2mm}
\caption{Experiment Testbed (Two physical machines were used and indicated in green and orange boxes.)}
\label{fig:Testbed}
\vspace{-2mm}
\end{figure}

\fi

\vspace{-1mm}
\subsection{Experimental Setup}
Our evaluation setup is depicted in Figure~\ref{fig:Testbed}. We used two PCs with Ubuntu 20.04 to deploy our testbed. The PC labeled in the green box is responsible for running the 5G core network, SDN controller and network monitoring tool. The PC labeled in the orange box is responsible for emulating the radio access network including the gNB and the user equipment using UERANSIM~\cite{ueransim}. We used Free5GC~\cite{free5gc} as our 5G core network and implemented our DANSM middleware within the Ryu controller~\cite{RYU} as part of the Free5GC control plane. We deployed all the 5G core functions inside Docker containers and orchestrated all network functions using Docker Compose. 

Validation using emulation of the factory floor including its 5G radio network and the robotic arms representing the UEs is justified because DANSM focuses on alleviating the bottlenecks in the 5G core, and moreover, conducting experiments on operational factory floors is hard unless there is a dedicated testbed for such a purpose.

\ifshort  
All the network traffic within our testbed is routed using Open vSwitch~\cite{openvswitch} and monitored by sFlow-RT~\cite{sFlow}. 
The traffic from UERANSIM is generated using iPerf3~\cite{iPerf}. All the UEs use TCP for guaranteeing communication reliability, and all the UE request rates follow the Poisson Distribution. We evaluate the efficacy of DANSM's heuristics algorithm that solves the optimization problem detailed in~\cite{Min2021DANSM}. 
\else 
All the network traffic within our testbed is routed using Open vSwitch~\cite{openvswitch} and monitored by sFlow-RT~\cite{sFlow}. 
The traffic from UERANSIM is generated using iPerf3~\cite{iPerf}. All the UEs use TCP for guaranteeing communication reliability, and all the UE request rates follow the Poisson Distribution. We evaluated DANSM in the application plane using the metrics to solve the optimization goal that are defined in Eq.(9). 

\vspace{-1mm}
\subsection{Baseline Algorithms}
We compared DANSM with the Modified Greedy Algorithm (MGA), which is a heuristic algorithm we developed in prior work and had applied to the dynamic switch migration problem~\cite{9659533}. MGA aims to minimize the switch queuing latency and the controller processing latency. Its objective function targets minimizing the average load of SDN controllers and the switch migration cost under dynamic traffic change. 

We also compared DANSM with conventional bin packing algorithms: First Fit Descending (FFD) and Best Fit Descending (BFD) algorithms~\cite{johnson1973near}. In our case, the FFD algorithm  starts with sorting the UEs in the UE freelist in descending order based on the UE priority. For each UE, after assigning it to the matched network slice (NS), FFD will scan the UPFs within the matched network slice in order and assign the current UE to the first UPF that is able to process the traffic from the current UE.  Similar to FFD, BFD will first sort all the UEs in the UE freelist in descending order based on the UE priority. For each UE, after assigning it to the matched network slice, the algorithm will scan all the UPFs within the matched network slices and assign the UE to a UPF where it fits the tightest.

\vspace{-1mm}
\subsection{Evaluating Load Balancing for Data Plane}

Load balancing is a significant objective for all the dynamic scheduling algorithms we compared. The unbalanced load in the 5G data plane will lead to unexpected queuing time on the UPF side and therefore increase the UPF processing time and hence the overall latency, which hurts system performance. To evaluate the performance of all the algorithms, we configured 10 UPFs in the data plane for all the NSs and set up 16 UEs to generate network traffic. The task priority dynamically decides the number of UPFs within the NS. We run each algorithm for 25 mins (1500 secs) for the sake of illustration; a 25 mins duration is long enough for the system to reach stability and the metrics fluctuate within about $10\%$.  Moreover, the UE request rates follow the same Poisson Distribution in every algorithm.

Figure~\ref{fig:DataPlaneLoad} uses the mean and standard deviation metrics to evaluate the UPF loads in the data plane. The x-axis indicates the time, and the y-axis shows the UPF loads in the system (We only calculated the UPF in use, not all the UPFs in the UPF Pool). From 0 mins to 5 mins, the system is in the warm-up status, and the UE containers are built up, registered to the 5G core network in succession and randomly assigned to the UPFs. Then, after capturing a number of connected and available UEs, the scheduling algorithms start to work. After 15 mins, the experimental results show that all the algorithms have a similar standard deviation, which indicates that the connections of all the UEs are stable while running the dynamic scheduling algorithm. 
However, our DANSM has a better mean value and a better standard deviation as well after 19 mins under dynamic network traffic, which indicates that DANSM is able to handle the extreme case, where all the UE loads are assigned to a few UPFs. MGA aims at minimizing both mean load and the standard deviation of the UPFs; therefore it can address the extreme case as well. Moreover, the FFD and the BFD have the nature of packing the load with the minimum resources; therefore they may lead to the extreme case occasionally. At 25 mins, the UPF average load of DANSM is $70\%$ of BFD, $91\%$ of MGA, and $93\%$ of FFD. In addition, the standard deviation of UPF load of BFD is 1.28 times more than DANSM. To sum up, DANSM efficiently and effectively achieves the optimization objective and significantly balances the UPF loads thereby minimizing the queuing latency and improving the performance of the adaptive robotic repair system.  

\begin{figure}[htb]
\ifshort
\vspace{-2mm}
\else
\vspace{-2mm}
\fi
\centerline{\includegraphics[width=0.99\columnwidth,keepaspectratio]{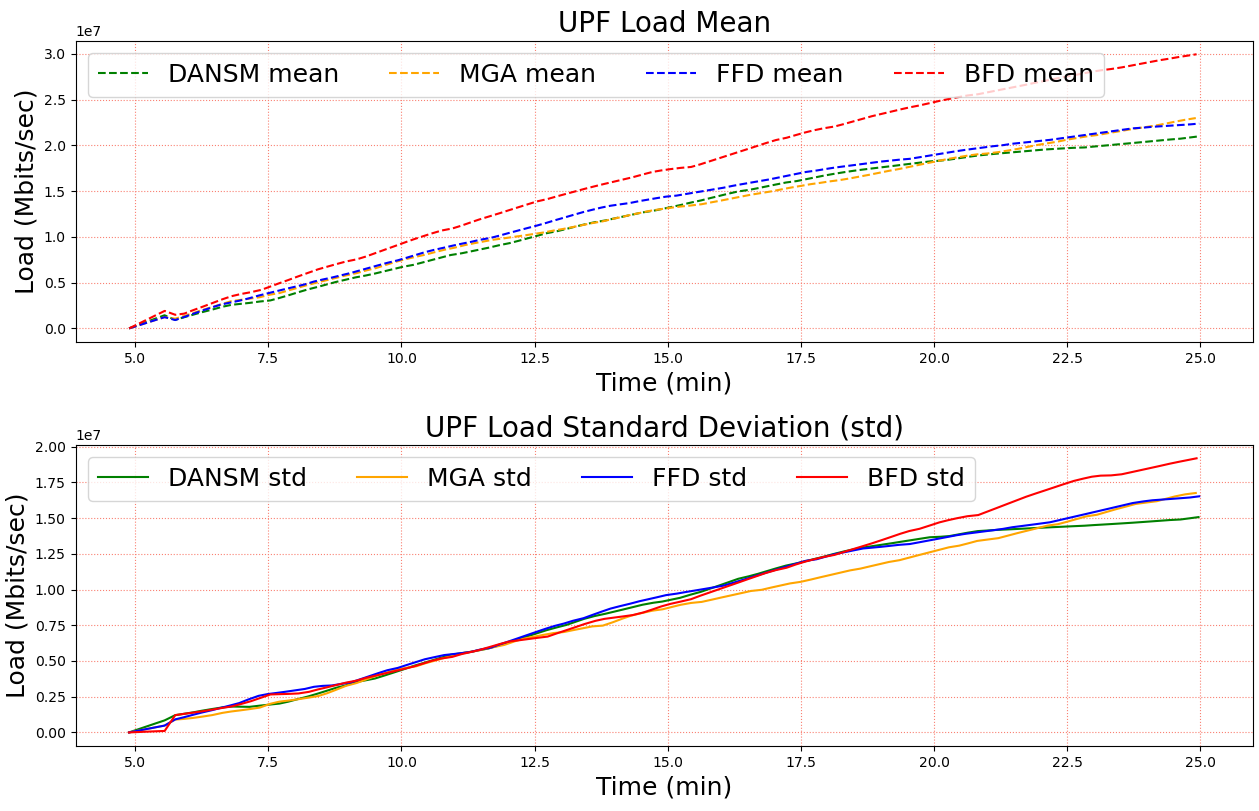}}
\vspace{-2mm}
\caption{Data Plane Load Mean and Standard Deviation(std)}
\label{fig:DataPlaneLoad}
\vspace{-2mm}
\end{figure}

\subsection{Evaluating End-to-End Response Time and Algorithm Efficiency}
We used iPerf3 to measure end-to-end response time and the number of tasks executed in the 25mins to evaluate the real-time performance of all the algorithms under dynamic network traffic. The evaluation result is shown in Figure~\ref{fig:TimeSpeed} and Figure~\ref{fig:NumOfTasks}. In Figure~\ref{fig:TimeSpeed}, the x-axis indicates the subtask names and the y-axis indicates the average response time for TCP transmission. In Figure~\ref{fig:NumOfTasks}, the x-axis shows the time and the y-axis shows the number of subtasks executed in the 25 mins. The results indicate that our DANSM approach outperforms MGA, FFD and BFD in both the time spent on the sub-tasks and completing the repair task. Although FFD shows a similar time in terms of a complete repairing task, DANSM performs better on the task with higher priority. Moreover, Figure~\ref{fig:NumOfTasks} shows that DANSM finished $34\%$ 
more subtasks than MGA, $64\%$ more subtasks than FFD and $22\%$ more subtasks than BFD in 25 mins.  Compared to the other three baseline algorithms, DANSM can specifically minimize the end-to-end response time for the sub-task with higher task priority and efficiently schedule all the subtasks under the dynamic network traffic thereby guaranteeing the performance of the adaptive robotic repairs.
\begin{figure}[htb]
\vspace{-1mm}
\centerline{\includegraphics[width=0.99\columnwidth,height=3cm]{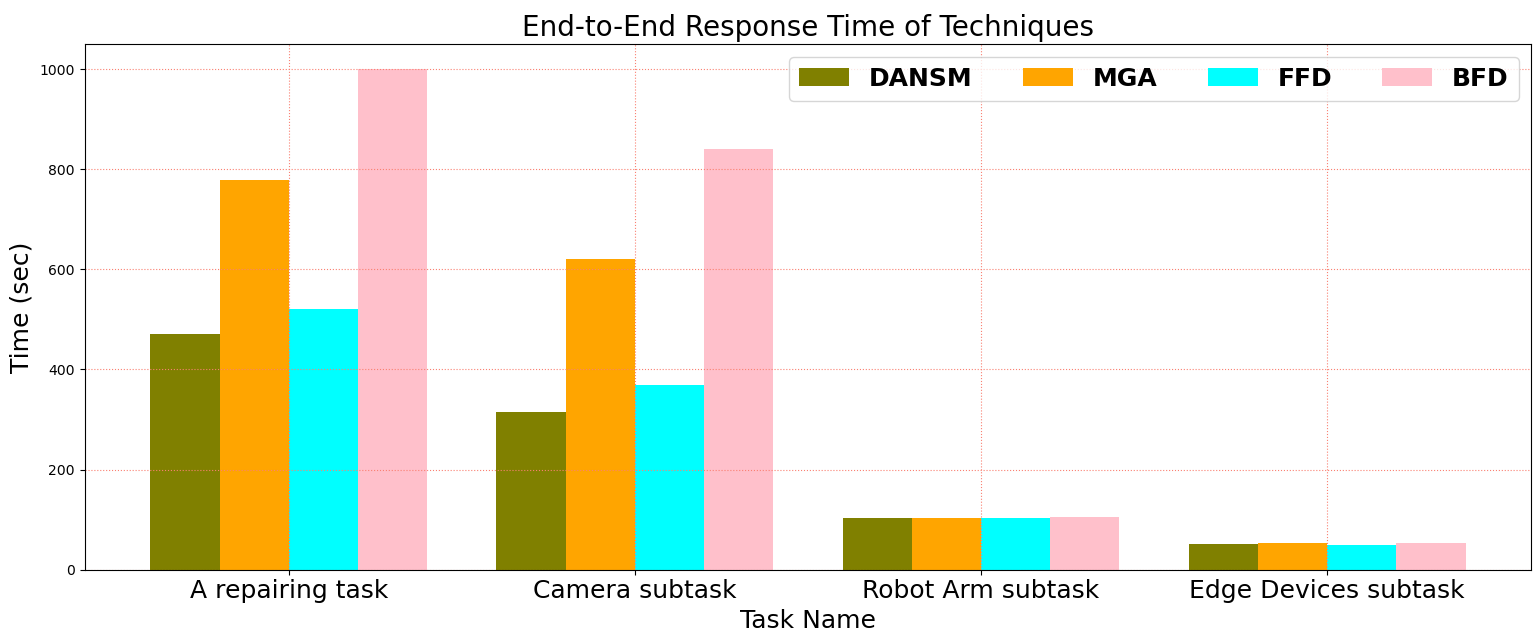}}
\vspace{-1mm}
\caption{End-to-End Response Time}
\label{fig:TimeSpeed}
\vspace{-1mm}
\end{figure}
\begin{figure}[htb]
\vspace{-1mm}
\centerline{\includegraphics[width=0.99\columnwidth,height=3.6cm]{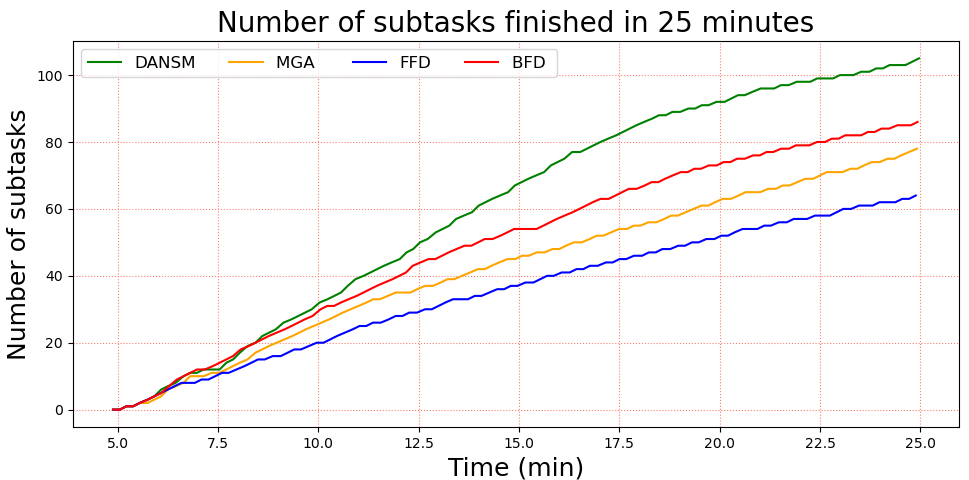}}
\vspace{-1mm}
\caption{Number of subtasks completing in 25 minutes}
\label{fig:NumOfTasks}
\vspace{-1mm}
\end{figure}

\vspace{-1mm}
\subsection{Summary}

Overall, the experimental results show that DANSM outperforms all the algorithms in both load balancing and end-to-end response time. The task priority and UE priority mechanism applied in DANSM significantly improved both sub-task and task completion performance. The multiple M/M/1 queuing models efficiently distributed the data plane traffic, thereby minimizing both queuing latency and propagation delay, and reducing the end-to-end latency that is critical in real-time industrial settings.

\section{Conclusion and Future Work}
\label{sec:ConclusionAndFutureWork}

\ifshort  

This paper presented DANSM, which is a middleware that executes as a
containerized microservice in the control plane of the 5G core.  DANSM offers
dynamic and autonomous network slice management to meet the QoS needs of IIoT
applications.  The paper demonstrated DANSM's capabilities for a 5G-based
adaptive robotic repair use case from the manufacturing domain.  The key
contributions in DANSM include a sub-task priority allocation  algorithm, a
queuing theory-based optimization problem formulation to minimize queuing
delays and improve latencies, and a heuristics approach to solve this
optimization problem at runtime. 

Our future work in this area is seeking to validate our approach on other IIoT
use cases and at larger scales. Presently, DANSM manages network slices at the
5G core and future work will also look at end-to-end slices, e.g., including
RAN-based slices.  Incorporating resilience and availability as additional QoS
dimensions is also part of our future work.

Software artifacts used in this research are available at \url{https://github.com/minziran/DANSM}

\else     
This paper presented DANSM, which is a software-defined solution that executes as a
containerized microservice in the control plane of the 5G core.  DANSM offers
dynamic and autonomous network slice management to meet the QoS needs of IIoT
applications.  The paper demonstrated DANSM's capabilities for a 5G-based
adaptive robotic repair use case from the manufacturing domain.  The key
contributions in DANSM include a sub-task priority allocation  algorithm, a
queuing theory-based optimization problem formulation to minimize queuing
delays and improve latencies, and a heuristics approach to solve this
optimization problem at runtime. 

Our future work in this area is seeking to validate our approach on other IIoT
use cases with real-time data and at larger scales. Presently, DANSM manages network slices at the 5G core only and future work will also look at end-to-end slices, e.g., including
RAN-based slices.  Incorporating resilience and availability as additional QoS
dimensions is also part of our future work.

\fi       

\balance

\bibliographystyle{ieeetr}
\bibliography{references}

\end{document}
\endinput